\documentstyle[11pt,epsfig]{article}
\input mssymb
\newcommand{\psbild}[5]
{\par
 \begin{figure}[#1]
 \begin{center}
 \begin{minipage}{0cm} \end{minipage}
 \begin{minipage}{#3}
 \refstepcounter{figure}\label{#2}
 \epsfxsize=#3
 \epsffile{#4}
 \end{minipage}
 \end{center}
 \hfill
 \begin{minipage}{0cm} \end{minipage}
 \begin{center}
 \parbox{15cm}{\baselineskip11pt{\rm Figure {#2}}: {#5}}
 \end{center}
 \end{figure}}

\newcommand{\beq}{\begin{equation}}
\newcommand{\eeq}{\end{equation}}
\newcommand{\bdm}{\begin{displaymath}}
\newcommand{\edm}{\end{displaymath}}
\newcommand{\bea}{\begin{eqnarray}}
\newcommand{\eea}{\end{eqnarray}}

\begin{document}
\setcounter{page}{0}
\topmargin 0pt
\newpage
\setcounter{page}{0}
\begin{titlepage}
\vspace{0.5cm}
\begin{center}
{\Large {\bf The generic soliton of the $A_n$ affine Toda field theories}} \\
\vspace{1cm}
{\large Edwin J. Beggs$^*$ and Peter R. Johnson$^\dagger$}\\
\vspace{1.0cm}
{\em $^*$Department of Mathematics,\\
University of Wales at Swansea,\\
Singleton Park,\\
Swansea,\\
 SA2 8PP,UK.} \\ 
\vspace{0.2 cm}
{\em $^\dagger$Department of Physics,\\
University of Wales at Swansea,\\
Singleton Park,\\
Swansea,\\
 SA2 8PP,UK.} \\ \end{center}
\begin{center}{\bf Abstract} \\ \end{center} 

In this note we show that the single soliton solutions known
previously in the $1+1$ dimensional affine Toda field theories from a
variety of different methods \cite{H1,MM,OTUa,OTUb}, are in fact not
the most general single soliton solutions. We exhibit single soliton
solutions with additional small parameters which reduce to the
previously known solutions when these extra parameters are set to
zero. The new solution has the same mass and topological charges as
the standard solution when these parameters are set to zero. However
we cannot yet completely rule out the possibility that other solutions
with larger values of these extra parameters are non-singular, in the
cases where the number of extra parameters is greater than one, and if
so their topological charges would most likely be different. 
\end{titlepage}\baselineskip=17pt 
\section{Introduction}
The affine Toda field theories have equations of motion based on the
root system of an affine algebra $\hat{g}$. Let $\alpha_i, i=1,\ldots,
r$ be the simple roots of the Lie algebra $g$, and $\alpha_0$ be minus
the highest root $-\psi$, then the equations of motion of the affine
Toda field theories, for $\phi(x,t)$ an $r$-component scalar field,
are \beq
{\partial^2\phi\over\partial t^2}-{\partial^2\phi\over\partial x^2}
+{4\mu^2\over\beta}\sum_{i=1}^r
m_i{\alpha_i\over\alpha_i^2}e^{\beta\alpha_i.\phi} =0.\label{eq: Toda}\eeq
Here $m_i$ are certain integers such that 
$${\psi\over\psi^2}=\sum_{i=1}^r m_i{\alpha_i\over\alpha_i^2},$$
and $m_0=1$, so that $\sum_{i=0}^r m_i{\alpha_i\over\alpha_i^2}=0$. If
$\beta$ is purely imaginary then we see that a constant $\phi$ such
that $\alpha_i\cdot\phi\in {2\pi\over |\beta|}{\Bbb Z}$ is a
solution. These constant values make up the weight lattice of the
coroot algebra $\Lambda_{W}(g^{\vee})$, $\phi\in {2\pi\over
|\beta|}\Lambda_{W}(g^{\vee})$, and these values lie at the degenerate
minima of the potential in the Lagrangian to the affine Toda
model. Hence we expect soliton solutions to exist interpolating
different values of these minima at $x\rightarrow \infty$ and
$x\rightarrow -\infty$. Some solutions were first found by Hollowood
\cite{H1} for the $A_n$ theories using the Hirota method, and some other
solutions by MacKay and McGhee \cite{MM} for other theories also using
the Hirota method.
This method however is somewhat unsatisfactory and a more
powerful method exploiting the representation theory of affine
Kac-Moody algebras, and the vertex operators, was developed for all
simply-laced algebras in
\cite{OTUa, OTUb}. These methods agreed on what the solutions were for
the $A_n$ theories, it was thought that the single soliton
solutions were, for the species $i=1,\ldots,r$ of soliton, associated
with a node on the Dynkin diagram, and $\lambda_j$ a fundamental
weight of $g$,
\beq
e^{-\beta\lambda_j\cdot\phi}={1+Q_1\omega^{ij}W_i\over
1+Q_1W_i}\label{eq: Ansol}, \eeq
where \beq\omega=e^{2\pi i\over h},\qquad
W_i(\theta)=e^{m_i(e^{-\theta}x_+-e^{\theta}x_-)},\qquad x_\pm=t\pm x
\label{eq: W}\eeq
$$h=n+1,\qquad m_i=2\mu\sin\bigl({\pi i\over h}\bigr),\qquad Q_1\in {\Bbb C},$$
and $\theta$ is the rapidity of the soliton. The position of the
soliton is proportional to $\log|Q_1|$. The
solution (\ref{eq: Ansol}) is also singular at particular values of
the phase of $Q_1$ given by the zeroes of either the numerator or the
denominator of (\ref{eq: Ansol}). In this case the singular values of
$Q_1$ are simply given by straight lines joining the origin and
extending to infinity. For simplicity, we restrict our attention to
$h=4$, and $i=1$, then the singularities in the $Q_1$ plane are with
phases
 $0,\frac\pi 2,\pi,\frac{3\pi}2$.

The topological charge of the soliton is defined as
$$T={|\beta|\over 2\pi}\Bigl(\phi(\infty,t)-\phi(-\infty,t)\Bigr)\in
\Lambda_W(g^\vee),$$
and is independent of time $t$. The topological charge is a continuous
function of $Q_1$ and takes discrete values, so it will be constant as
we vary $Q_1$ within each of the four regions, but will not be
defined if $Q_1$ touches the singular lines. In fact we expect the
charge to jump and take a different value as $Q_1$ moves between the
regions  separated by the singularities. This is
corroborated in detail by work done by McGhee \cite{McGhee} for the $A_n$
theories, where the charges are computed. For the remaining Toda
theories, and the previously known single soliton solutions, the
singular regions in the $Q$ plane are also given by straight lines
joining the origin and extending to infinity, but the analysis is
slightly more complicated because more than one power in $QW$ is
present in the numerator and denominator of
$e^{-\beta\lambda_j\cdot\phi}$, compare with the formula (\ref{eq:
Ansol}) for $A_n$. We shall see that this will not be the case in
general for even the simpler $A_n$ theories. 

These methods \cite{H1,OTUa,OTUb} 
also agreed on a form for the two-soliton solution. Here we follow 
\cite{OTUa,OTUb}, where the two-soliton solution is understood in terms of a
special function $X^{jk}(\theta_j-\theta_k)$, which is a function
obtained when we normal order the two vertex operators
$F^{j}(\theta_j)$ and $F^{k}(\theta_k)$ associated with the solitons
of species $j$ and $k$ in the two-soliton solution. Here $\theta_j$
and $\theta_k$ are the rapidities of the two solitons which must be
real in order to make physical sense.
$$F^{j}(\theta_j)F^{k}(\theta_k)=X^{jk}(\theta_j-\theta_k)
:F^{j}(\theta_j)F^{k}(\theta_k):$$
and $X^{jk}(\theta)$ can be given explicitly by 
\beq X^{jk}(\theta)=\prod_{p=1}^h\Bigl(1-e^\theta e^{{\pi i\over
h}(2p+{c(j)-c(k)\over 2})}\Bigr)^{\gamma_j\cdot\sigma^p\gamma_k}.\label{eq:
X} \eeq
Here $c(j)=\pm 1$ is a particular `colour' depending on a
bi-colouration of the Dynkin diagram of $g$, where the soliton of
species $j$ is associated with a node of the Dynkin
diagram. Also $\gamma_j=c(j)\alpha_j$, and $\sigma$ is a special element of
the Weyl group known as the Coxeter element \cite{OTUa,OTUb}. For the
$A_n$ theories the two-soliton solution (species $j$ and species $k$)
is
\beq e^{-\beta\lambda_i\cdot\phi}={1+Q_1\omega^{ik}W_k+Q_2\omega^{ij}W_j+
X^{kj}(\theta_k-\theta_j)Q_1Q_2\omega^{i(k+j)}W_kW_j\over
1+Q_1W_k+Q_2W_j+
X^{kj}(\theta_k-\theta_j)Q_1Q_2W_kW_j}\label{eq: 2sol}\eeq
The coefficient $X^{kj}(\theta)$ tells us a surprising amount about
the interaction of two solitons, albeit where the single solitons are
the ones given by (\ref{eq: Ansol}) and not the more general ones
which we are about to discuss. The time delay experienced by the
soliton $k$ as it interacts with soliton $j$ is proportional to $\log
X^{kj}(\theta)$ \cite{FJKO}. $X^{kj}(\theta)$ can also be extrapolated
to the exact S-matrix of the solitons \cite{PRJ}.

In this note we are not directly concerned with these properties of
$X^{kj}(\theta)$ related to the interaction of two
solitons, because we shall take the case where $X^{kj}(\theta)$
vanishes. This has not been treated in the literature before because 
it was previously thought that (\ref{eq: 2sol}) did not have
real total energy and momentum when $\theta_k-\theta_j$ is at the
zeroes of $X^{kj}(\theta)$. It was also thought in \cite{MIH} that the
restricted solution was singular. Some of these solutions were
mentioned by Caldi and Zhu \cite{ZC}, but not properly identified as 
true single solitons and also not fully discussed. 
We shall see how the restriction
works in the next section.  However
before we do this we shall briefly discuss an alternative scheme
developed by us \cite{BJ,PRJ}, different from \cite{H1,OTUa,OTUb}, for
finding soliton solutions, which for the moment is restricted to the
$A_n$ theories. This method is based on the inverse scattering method
\cite{Edwin,Novikov}. Now the integrability of the affine Toda systems
follows from the zero-curvature condition
\beq
[\partial_++A^+,\partial_-+A^-]=0\label{eq: zcc},\eeq
where $A^{\pm}$ is given by
\beq
A^{\pm}\ =\ \pm\frac12\beta\partial_\pm(\phi.H)\ \pm\ \lambda^{\pm 1}\mu
e^{\pm\frac12\beta\phi.H}E_{\pm 1}e^{\mp\frac12\beta\phi.H},
\label{eq: As} \eeq
and where $H$ is the Cartan-subalgebra of $g$, and
$E_{+1}={E_{-1}}^\dagger=\sum_{i=0}^r\sqrt{m_i}E_{\alpha_i}\in g$, for
$E_\alpha$ the step operator in $g$ corresponding to the root
$\alpha$. It is easy to see that $[E_{+1},E_{-1}]=0$.
The compatibility condition
$\partial_+\partial_-\Phi=\partial_-\partial_+\Phi$ for $\Phi$ the
solution to the linear system
\beq \partial_\pm\Phi=A^\pm\Phi, \label{eq: lin_sys}\eeq
implies the zero-curvature condition (\ref{eq: zcc}), and here
$\Phi(x,t,\lambda)$ is valued in the loop group $\hat{G}$ of $\hat{g}$
\cite{PS}. Therefore any solution $\Phi$ to (\ref{eq: lin_sys}) 
in $\hat{G}$ will generate a solution to the affine Toda field
equations of motion (\ref{eq: Toda}). It turns out however that
solutions $\Phi(\lambda)$ which are analytic in $\lambda$ (except for some
essential singularities at $\lambda=0, \infty$, which can always be
subtracted off) generate the trivial solution $\phi=0$ to (\ref{eq:
Toda}). It is the solutions with poles which precisely generate the
soliton solutions. The number of poles present (modulo a discrete
rotational symmetry in $\lambda$) gives the number of solitons
generated, and the residues of the poles determine the species and
positions, and the other degrees of freedom.

We can exhibit the solutions $\Phi(\lambda,x,t)$ which generate a
single soliton explicitly, these have a pole at
$\lambda=\omega\alpha$, for some $\alpha\in {\Bbb C}$. However there
is also a discrete rotational symmetry $U\Phi(\lambda)U^\dagger=f(\lambda)
\Phi(\omega\lambda)$, where $f(\lambda)$ is a scalar function,
for some matrix $U$ which commutes with $H$, and
$U^{h}=1$,
this follows from the linear system (\ref{eq: lin_sys}) and the
property $UE_{\pm 1}U^\dagger=\omega^{\pm 1}E_{\pm 1}$. Therefore
there must also be poles in $\Phi(\lambda,x,t)$ at 
$\lambda=\omega\alpha,\omega^2\alpha,\ldots,\omega^{h-1}\alpha,\alpha$. 
The general
solution for $\Phi(\lambda)$ incorporating this symmetry is\newpage
\bea
\Phi(\lambda)=\Bigl(P+UPU^\dagger\Bigl({\lambda-\alpha\over\lambda-
\omega\alpha}\Bigr)
&+&U^2P{U^\dagger}^2
\Bigl({\lambda-\alpha\over\lambda-\omega^2\alpha}\Bigr)+ \cdots \cr\cr\cr
&+&U^{h-1}P{U^\dagger}^{h-1})\Bigl({\lambda-\alpha\over\lambda-\omega^{h-1}
\alpha}\Bigr)\Bigr)e^{-{\beta\phi\cdot
H}/2}
\label{eq: Phi}, \eea
where $P$ is the unique matrix valued projection such that 
\beq
P+UPU^\dagger + \cdots + U^{h-1}P{U^\dagger}^{h-1}=1,\qquad
PUP=PU^2P=\cdots=PU^{h-1}P=0\label{eq: conditions},\eeq
as explained in \cite{BJ,PRJ}.
Multi-soliton solutions are generated by multiplying these (\ref{eq:
Phi}) together in
the loop group $\hat{G}$. 

The space-time dependence of the projection can be easily found
\cite{BJ,PRJ}, it is the unique projection which projects onto
the space 
\beq
V=e^{-\mu(\omega\alpha
E_{+1}x_+-\omega^{-1}\alpha^{-1}E_{-1}x_-)}V_0\label{eq: evol},\eeq
where $V_0$ is some initial arbitary one-dimensional space, and which
satisfies the conditions (\ref{eq: conditions}).
In a basis where $H$ is diagonal, 
\beq
E_{+1}=\pmatrix{0 & 1 & 0 & \cdots & & \cr & 0 & 1 & 0 & \cdots & \cr
& & 0 & 1 & 0 & \cdots \cr \vdots & & &\ddots & \ddots & \cr
0 & 0 & \cdots & &0& 1 \cr
1 & 0 & \cdots & & & 0 }\quad,\quad 
E_{-1}=E_{+1}^\dagger=\pmatrix{0 & 0 & \cdots &  & & 1\cr
1& 0 & \cdots & & & \cr
0& 1& 0 & \cdots & & \cr 0& 0& 1 &0 &\cdots & \cr
\vdots & & &\ddots & \ddots & \cr
0 & \cdots &  & & 1& 0 },
\eeq
the eigenvectors of $E_{\pm 1}$ are 
$$v_0=\pmatrix{1\cr 1\cr 1 \cr \vdots \cr 1},
v_1=\pmatrix{1\cr \omega \cr \omega^2 \cr \vdots \cr \omega^{h-1}},
\cdots,
v_r=\pmatrix{1\cr \omega^r \cr \omega^{2r} \cr \vdots \cr \omega^{r(h-1)}},
\cdots,
v_{h-1}=\pmatrix{1\cr \omega^{-1} \cr \omega^{-2} \cr \vdots \cr 
\omega^{-(h-1)}}
$$
with eigenvalues 
$$E_{\pm 1}v_r=\omega^{\pm r}v_r,\quad r=0,\ldots,h-1.$$
If we write 
\beq V=<\pmatrix{1 \cr A_1 \cr A_2 \cr \vdots\cr A_{n}}>,\label{eq:
norm} \eeq then 
the soliton solutions are given by 
\beq e^{-\beta\lambda_i\cdot\phi}=A_i, \label{eq: sols} \eeq
as explained in \cite{BJ,PRJ}. So the choice of initial subspace $V_0$
gives us different single soliton solutions. We see that the choice 
\beq V_0=<v_0+Q_1v_j>\label{eq: first_choice} \eeq
gives us the one-soliton solution (\ref{eq: Ansol}) for a soliton of
species $j$, this follows because after normalizing the first
component of $V$ to agree with (\ref{eq: norm}), and solving for the
space-time dependence using (\ref{eq: evol}), we find that
$$A_i={1+Q_1\omega^{ij}W_j\over 1+Q_1W_j}.$$
However we also see that the choice 
\beq V_0=<v_0+Q_1v_k+Q_2v_j>\label{eq: second_choice}\eeq 
is just as good as the first choice (\ref{eq: first_choice}), it is
just a more general one-dimensional subspace than (\ref{eq:
first_choice}). We shall see that this solution is actually a
restriction of the two-soliton solution (of species $j$ and species
$k$) such that $X^{kj}(\theta)=0$, (compare with equation (\ref{eq:
2sol})), but more importantly its energy is real and it has the same
mass as the soliton of species $j$. Furthermore, provided $Q_1$ is sufficiently
small when compared with $Q_2$, it is non-singular for all $x$ and $t$ and
is therefore a bona-fide solution.

\section{The more general one-soliton solution}
\subsection{The two parameter case}
\setcounter{equation}{0}
The solution given by the initial subspace (\ref{eq: second_choice})
is from (\ref{eq: evol}), (\ref{eq: norm}) and (\ref{eq: sols})
\beq
e^{-\beta\lambda_i\cdot\phi}={1+Q_1\omega^{ik}U+Q_2\omega^{ij}W_j(\theta_j)
\over 1+Q_1U+Q_2W_j(\theta_j)}, \label{eq: new_sol} \eeq
where $W_j(\theta)$ is given by equation (\ref{eq: W}), and we have
chosen the phase of $\omega\alpha$ so that $\omega\alpha
=i\omega^{-j/2}e^{-\theta}$, for $\theta$ a real rapidity. Then we have
\bea
U&=&e^{\mu2\sin\bigl({k\pi\over h}\bigr)(e^{{\pi i(j-k)\over
h}}e^{-\theta}x_+-e^{-{\pi i(j-k)\over
h}}e^{\theta}x_-)}\cr \cr
&=&W_k\Bigl(\theta-{\pi i(j-k)\over
h}\Bigr) \label{eq: notatrest}. \eea
For simplicity we now put the soliton at rest by setting $\theta=0$, and
then
\bea W_j&=&e^{m_j (2x)} \cr
U&=&=e^{m_k\bigl(2\cos\bigl({(j-k)\pi\over
h}\bigr)x+i2\sin\bigl({(j-k)\pi\over h}\bigr)t\bigr)} \label{eq: atrest}
\eea
We can then perform a Lorentz transformation on this to get back to a
moving soliton with $\theta\neq0$.

We choose $j$ and $k$ so that 
$$m_j>m_k\cos\Bigl({(j-k)\pi\over
h}\Bigr),$$
hence for large $x$, $W_j$ will dominate  $U$ and
$e^{-\beta\lambda_i\cdot\phi}$ will have the same limit as when
$Q_1=0$. Since the mass of the soliton is only a function of the limit
of the solution $\phi$
when $x\rightarrow\pm\infty$ (the energy density can be written as a
total derivative \cite{OTUa}), and the limit as $x\rightarrow -\infty$
of $e^{-\beta\lambda_i\cdot\phi}$ is $1$ for all $Q_1\in {\Bbb C}$, we conclude
that the mass of the solution (\ref{eq: new_sol}) is real and the same
as for the case $Q_1=0$. This mass is 
$m_j=2\mu\sin\bigl({j\pi\over h}\bigr)$, see \cite{OTUa}. If we
re-insert the $\theta$ dependence into $W_j$ and $U$, then the total
energy and momentum $P^\pm$ in light-cone co-ordinates is 
$P^\pm=m_je^{\mp\theta}$. This shows that the phase of $\omega\alpha$
has been chosen correctly so that for $\theta$ real, we get real total
energy and momentum, and that $\theta$ agrees with the standard
definition of rapidity.

If we now take the two-soliton solution (\ref{eq: 2sol})
\beq e^{-\beta\lambda_i\cdot\phi}={1+Q_1\omega^{ik}W_k(\theta_k)+Q_2
\omega^{ij}W_j(\theta_j)+
X^{kj}(\theta_k-\theta_j)Q_1Q_2\omega^{i(k+j)}W_k(\theta_k)W_j(\theta_j)\over
1+Q_1W_k(\theta_k)+Q_2W_j(\theta_j)+
X^{kj}(\theta_k-\theta_j)Q_1Q_2W_k(\theta_k)W_j(\theta_j)}.
\label{eq: 2solb}\eeq
This equation makes physical sense for $\theta_k$ and $\theta_j$ real,
but will still satisfy the equations of motion (\ref{eq: Toda}) after
analytically continuing $\theta_k$ and $\theta_j$. In particular we
can set $\theta_k-\theta_j$ to be at a zero of
$X^{kj}(\theta_k-\theta_j)$. For the $A_n$ theories it is known
\cite{PRJ} from the formula (\ref{eq: X}) that this zero is at 
\beq \theta_k-\theta_j=i{\pi (k-j)\over h}.\label{eq: pos_zero} \eeq
Hence with this restriction, and with $\theta=\theta_j$, we recover the
formula (\ref{eq: new_sol}) from (\ref{eq: notatrest}) and (\ref{eq: 2solb}).
The energy and momentum $P^\pm$ of the generic two-soliton solution
(\ref{eq: 2solb}) is from \cite{OTUa}
\beq P^\pm=m_je^{\mp\theta_j}+m_ke^{\mp\theta_k}.\label{eq: two} \eeq
When evaluated at the analytically continued values (\ref{eq:
pos_zero}), this does not agree with our energy-momentum formula 
$P^\pm = m_je^{\mp\theta_j}$ derived for the
restricted solution (\ref{eq: new_sol}). This is somewhat surprising,
but a closer examination of the proof of the two-soliton result
(\ref{eq: two}) in \cite{OTUa}
requires $X^{kj}(\theta)\neq 0$, so that the dominant term for large
$x$ in both the
numerator and denominator of (\ref{eq: 2solb}) is $W_kW_j$.

Also note that, as remarked upon in \cite{FJKO},
 the case $j=k$ is empty, because $X^{jj}(0)=0$, and 
we recover the
standard one-soliton solution in the form
$$e^{-\beta\lambda_i\cdot\phi}={1+(Q_1+Q_2)\omega^{ij}W_j\over
1+(Q_1+Q_2)W_j}.$$

We now study the singularities of the solution (\ref{eq: new_sol}), in
the same way that we saw the singularities of (\ref{eq: Ansol}). For
illustrative purposes we restrict our attention to $h=4, j=1, k=3$.
With these values $m_j>m_k\cos\Bigl({(j-k)\pi\over h}\Bigr)$ (actually $j$
and $k$ are anti-solitons of each other, and $m_j=m_k$). Suppose that
$Q_2$ is given and that ${\rm Re}\ Q_2>0$, and ${\rm Im}\ Q_2<0$, say, 
for definiteness. We then sketch in the $Q_1$ plane the values where
the numerator and denominator of (\ref{eq: new_sol}) vanish, in the
rest frame of the soliton ($\theta=0$), as before. We also implicitly
absorb the time dependence $e^{im_k2\sin{(j-k)\pi\over h}t}$ from
(\ref{eq: atrest}) into $Q_1$.
%
%
\psbild{ht}{1}{10cm}{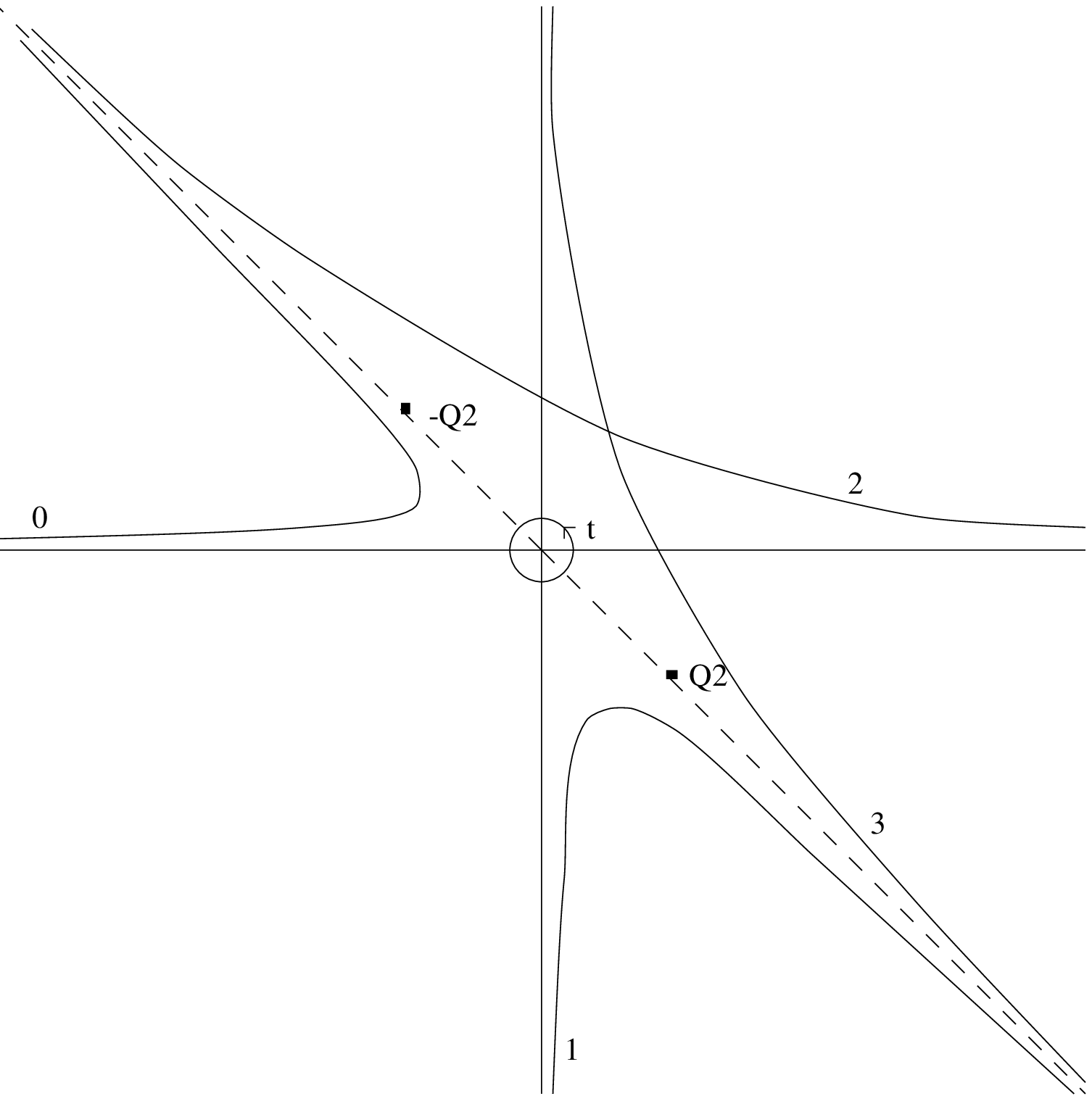}{Singularities
of (\ref{eq: new_sol}) in the $Q_1$ plane}
\noindent The label $0$ refers to the denominator of (\ref{eq: new_sol}), and
the labels $j=1,2,3$ refer to the numerator of $e^{-\beta\lambda_j\cdot\phi}$.

 We see that provided $|Q_1|$ is sufficiently small, a circle of
radius $|Q_1|$ centered at the origin will not intersect any of the
singular curves, and hence the solution (\ref{eq: new_sol}) is free of
singularities for all time $t$. As $t$ increases (in the rest frame)
we move round the circle as shown. This time dependence is interesting
because it shows that in the rest frame of the soliton, it is not
completely motionless, in constrast to the case with $Q_1=0$,
and that there is an incipient small beating motion,
similar to the way that a breather breathes. This is shown in Figure
5. Observe the difference between the simple step shown in Figure 4,
where $Q_1=0$, and the Figure 5.
The breather is an
analytic continuation of a soliton--anti-soliton solution
\cite{OTUb,MIH} which gives a real total energy and momentum, but
our solutions are certainly different from these breathers. Indeed in
\cite{MIH} it was thought that the analytic continuation of a two
soliton solution so that $X^{jk}(\theta)=0$ always gave a singular
solution, but the analysis in Figure 1 shows that this is not
necessarily the case.

Now Figure 1 was drawn given the assumption that $Q_2$ was known, and
that we had not accidently chosen a value which always gave a singular
solution. The phase of $Q_2$ is shown in the figure, and it is clear
by inspection that if we smoothly adjust the phase of $Q_2$ to either 
$0,\frac\pi 2,\pi,\frac{3\pi}2$, one of the four curves each in turn
will become a straight line passing through the origin, and the
solution must then be singular however small we choose $|Q_1|$. As the
phase of $Q_2$ becomes close to these four values we must choose
successively smaller values of $|Q_1|$ for the solution to be
non-singular. Hence in the $Q_2$ plane the singularities are the same
as the case with $Q_1=0$, but with the
understanding that $|Q_1|$ must be sufficiently small for a non-singular
solution. Also note that as $|Q_2|$ becomes smaller, but with the
phase of $Q_2$ fixed, the turning points of the curves in Figure 1 move
towards the origin, and tend to it in the limit when $|Q_2|\rightarrow
0$. Hence we must also adjust $|Q_1|$ depending on $|Q_2|$. 

We can also observe from Figure 1, that there are no new solutions for
any large values of $|Q_1|$, since the curves all go off to
infinity. A circle of large radius must intersect the curves.

Now a non-singular solution with appropiately small $|Q_1|$ is
continuously connected to the standard solution with $Q_1=0$,
therefore we
conclude from the continuity of the topological charge that the charge
is the same as the solution with $Q_1=0$. These are already well understood,
 at least for the $A_n$ theories, and have been calculated
\cite{McGhee}.
\subsection{More than two parameters}
We pick a soliton species $j$, and consider all soliton species $k$
such that 
\beq m_j>m_k\cos\Bigl({(j-k)\pi\over h}\Bigr)\label{eq:
mass_condition}.\eeq
Define the set $B_j$ to consist of all the integers $k$ such that
(\ref{eq: mass_condition}) holds. The
number of possibilities for $k$ will depend on $j$, but as we have
seen in the first case, we can always take the anti-soliton $k$ to the
soliton $j$ provided $j\neq k$. If the anti-soliton species is the
same as the soliton species, then $n$ must be odd, and
$k=\frac{n+1}2$. The property (\ref{eq: mass_condition}) must then be
true for all species $k$, since  the soliton $j$ is the heaviest. Also
note that for $j=1$ the only possibility for $k\in B_j$ is $k=n$.

We then choose the initial space
$$V_0=<v_0+\sum_{k\in B_j}Q_k v_k+Q_j v_j>$$
which generates the single soliton solution 
\beq
e^{-\beta\lambda_i\cdot\phi}={1+\sum_{k\in B_j}Q_k\omega^{ik}U_k+
Q_j\omega^{ij}W_j(\theta)\over
1+\sum_{k\in B_j}Q_kU_k+
Q_jW_j(\theta)}\label{eq: new_sol2}. \eeq
We have chosen the phase of $\omega\alpha$ exactly as before, in order
to get a purely exponential behaviour of $W_j$ in $x$ and $t$, namely
$\omega\alpha=i\omega^{-j/2}e^{-\theta}$, for $\theta$ real. Then 
\bea
U_k&=&e^{\mu2\sin\bigl({k\pi\over h}\bigr)(e^{{\pi i(j-k)\over
h}}e^{-\theta}x_+-e^{-{\pi i(j-k)\over
h}}e^{\theta}x_-)}\cr \cr
&=&W_k\Bigl(\theta-{\pi i(j-k)\over
h}\Bigr) \label{eq: notatrest2}, \eea
and $U_j=W_j$. The set $B_j$ has been defined so that $W_j$ dominates
for large $x$ over the purely exponential parts of $U_k$. This is
enough to guarantee that the mass of the solution (\ref{eq: new_sol2})
is $m_j$.

It is clear that we can also derive (\ref{eq: new_sol2}) from a
restriction of a multi-soliton solution. We take the multi-soliton
solution with solitons of species in $B_j$ and also of species $j$,
 and with separate rapidities
$\theta_k$, $k\in B_j$ and $\theta_j$.
 The multi-soliton solution is \cite{OTUb}
$$e^{-\beta\lambda_i\cdot\phi}={1+\sum_{k\in
B_j}Q_k\omega^{ik}W_k(\theta_k)+Q_j\omega^{ij}W_j(\theta_j)+\mbox{higher
terms}\over
1+\sum_{k\in B_j}Q_kW_k(\theta_k)+
Q_jW_j(\theta_j)+\mbox{higher terms}}.$$
The higher terms all involve more than one power of $W$. They are all
multiplied by products of $X$'s which vanish when we take 
\beq \theta_k-\theta_j=i{\pi(k-j)\over h}.\label{eq: dagger}\eeq
 For example, a term $W_{k_1}W_{k_2}$ for $k_1,k_2\in B_j$, is
multiplied by $X^{k_1 k_2}(\theta_{k_1}-\theta_{k_2})$, but from
(\ref{eq: dagger}) $\theta_{k_1}-\theta_{k_2}=i{\pi(k_1-k_2)\over
h}$, and $X^{k_1 k_2}(\theta_{k_1}-\theta_{k_2})$ has a zero at this
point. 

We can now discuss the structure of the singularities of (\ref{eq:
new_sol2}), after setting $\theta=0$. We pick an $r\in B_j$, and
assume that the remaining $Q_k, k\in B_j-\{r\}$ are given, but are very
small. We also assume that $Q_j$ is given and can be as large as we
please. As before, we absorb the time dependent phase of $U_r$ into
$Q_r$, but we cannot do this for the other phases.
We then sketch the singularities of (\ref{eq: new_sol2}). We first of
all sketch the curves for the situation $h=4, j=2, r=1$, with all
$Q_k=0$, and $k\in B_j-\{r\}$. This is done in Figure 2. Observe that the
singular phases of $Q_2$ are now $0$ and $\pi$. We now make 
$Q_k$, $k\in B_j-\{r\}$ non-zero, but very small. We fix time
$t$, and sketch the curves for all $x$. The curve will have the same
asymptotic behaviour as the previous case, but will have moved by a
small amount. In Figure 3, regions are  sketched approximately by
varying all $x$ and $t$. Hence the solution is singular at some 
$x$ and $t$ in the singular regions, but it may be possible for the
circle centered at the origin with the time dependent phase 
to avoid a singularity even though it may pass through a singular
region. In fact for sufficiently large $Q_k$, $k\in B_j-\{r\}$, the
singular regions may cover the origin, and in that case $|Q_1|$ would
have to be non-zero for there to be the possibility of a non-singular
solution. This situation is too complicated for us to say anything
definite, for the time being. On the other hand, we certainly avoid
all singularities if $|Q_r|$
is sufficiently small so that $Q_r$ does not enter the singular
regions.  It is also clear that for the cases
where $Q_k, k\in B_j$ are small, the singular
lines in the $Q_j$ plane are the same as for the naive solution with 
$Q_k=0, k\in B_j$. 
 
\psbild{ht}{2}{10cm}{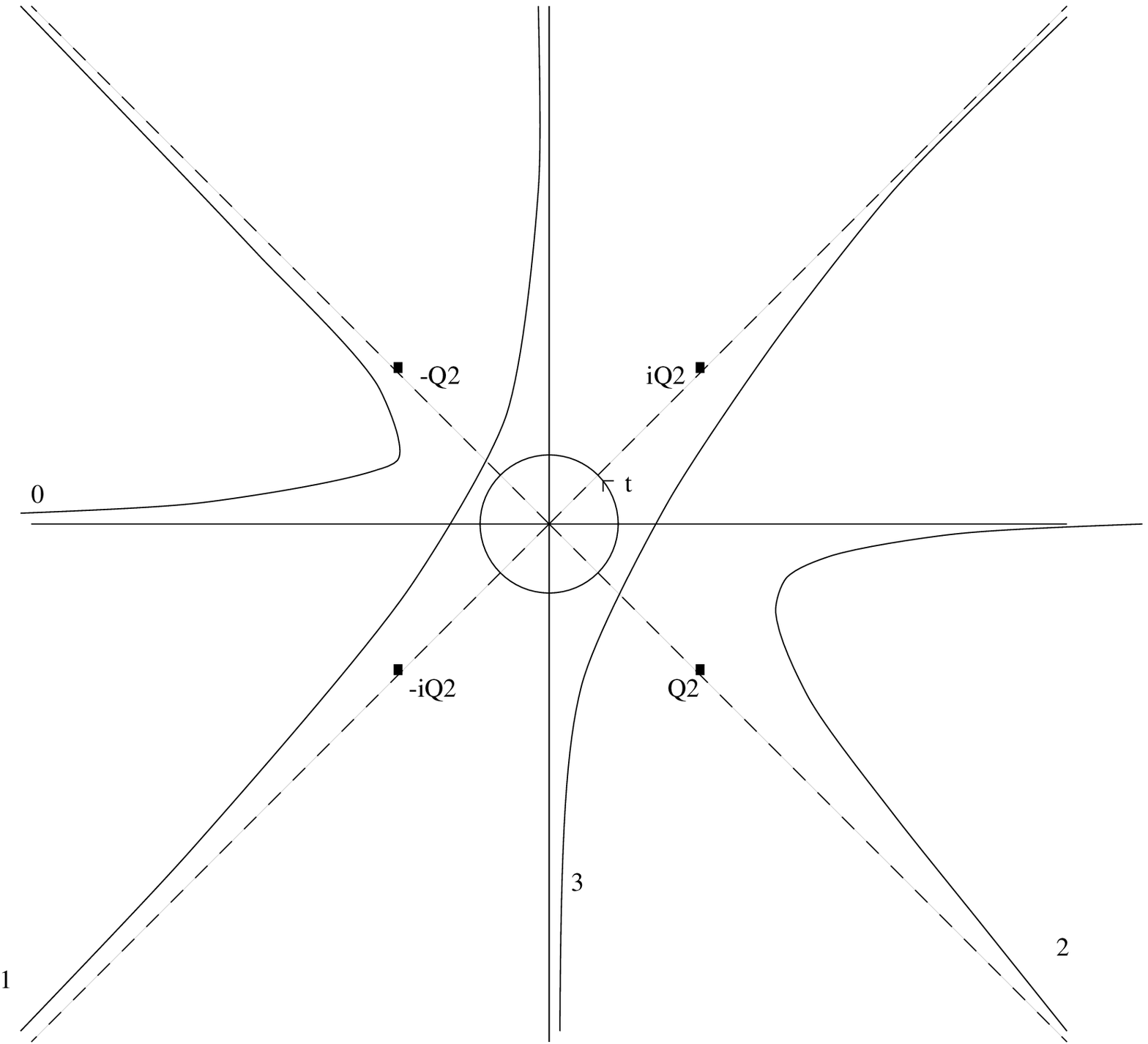}{Singularities
of $h=4, j=2, r=1$, in the $Q_1$ plane}
\psbild{ht}{3}{10cm}{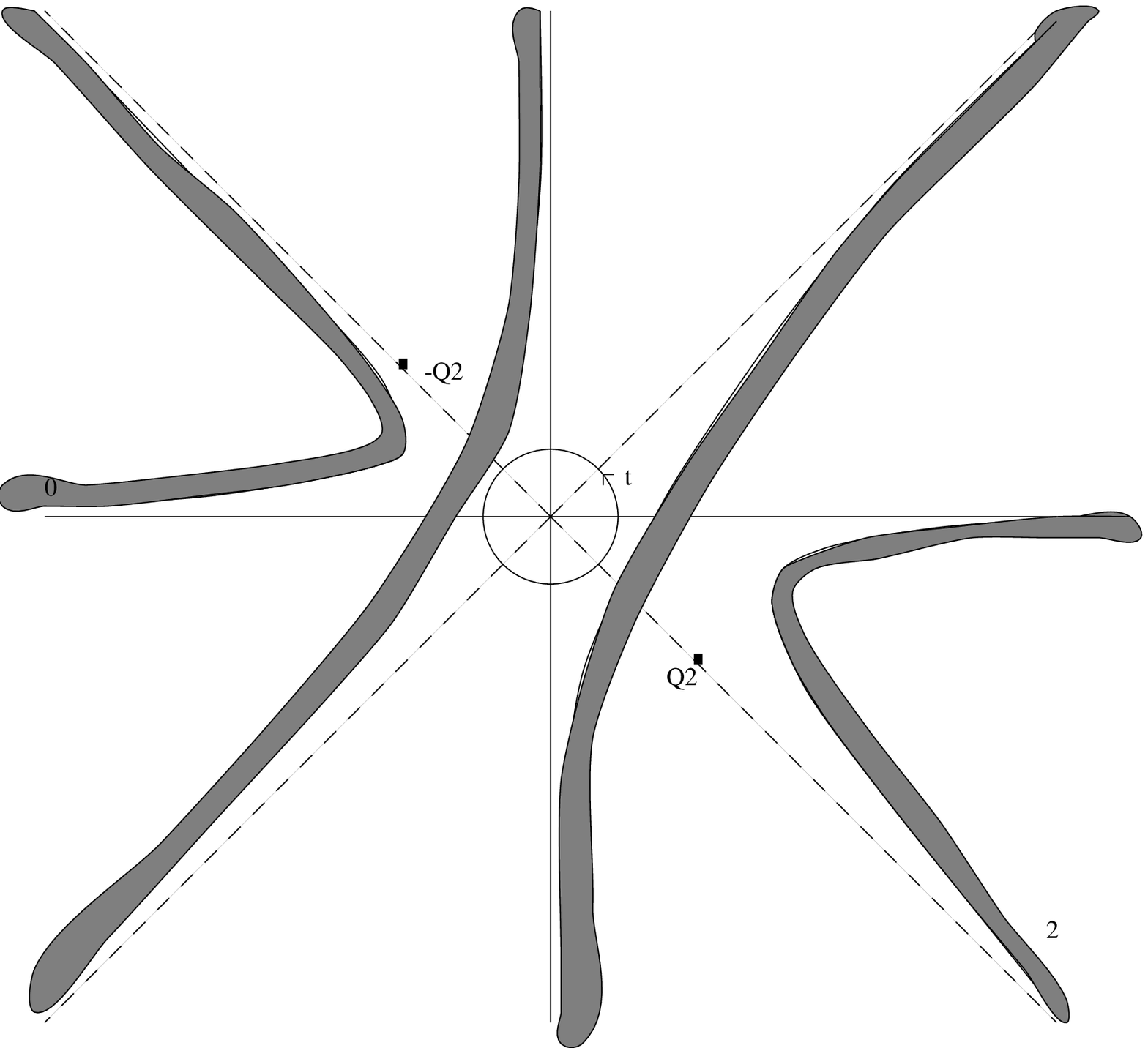}{Singularities
of $h=4, j=2, r=1, k=3$, with $Q_3\neq 0$ in the $Q_1$ plane}
\section{Discussion and Conclusions}
We have shown that the single soliton solutions in the affine Toda
field theories are not as simple as first thought, but that the
masses and topological charges of the new solutions with additional
small parameters are the same as the previous solutions obtained by
setting these extra parameters to zero. The inverse scattering method
is seen to be superior to the other methods, from the point of view of
being able to generate all single solitonic objects from a single
simple pole factor (\ref{eq: Phi}). It is not so obvious from the
restriction of the two-soliton solution, that the solutions that we
have derived are single soliton solutions. We also remark that
constructions of new solutions will follow in exactly the same way 
for the remaining affine
Toda theories, which we do not discuss. 

It is worth
considering these new classical solutions in the context of the
quantum theory. There are two issues here. The first is that
semi-classical techniques to finding quantum corrections to masses of
single solitons in \cite{Delius} and also in \cite{WM} only considered
the naive single soliton solutions. It is most likely that the
fluctuations indicated by the extra parameters introduce extra
corrections, and thus may invalidate the work in \cite{Delius} and
\cite{WM}.
The second point is that the circular degrees of freedom given by the
phases of the extra small parameters should lead to an excited set of
states all related to the same classical solution, when
quantized. Such circular degrees of freedom do not exist in the naive
solutions. One of us has already found exact S-matrices for the
solitons in simply-laced affine Toda field theories \cite{PRJ}, and
there it was discovered that associated with a pole due to the fusing
of solitons there was a string of further poles consistent with the
intepretation that they represent excited soliton states, with
increasing masses. Hopefully these new single soliton 
solutions will shed light on
the excited states seen in the S-matrix. It should be possible, for
example, to compute their masses and to compare them with the
positions of the poles of the string in the S-matrix.

We also remark that we have failed to find new solutions with
different topological charges from the single soliton solutions
already known, although we have not completely ruled out the
possibility of their existence, because of the different rates of
rotation of $U_k$, and therefore the possibility of non-singular
solutions even when the origin is contained within a singular region,
or the circle in the $Q_1$ plane passes through a singular region in
Figure 3.
 This was our intention when we first looked at the
single soliton solutions in more detail. If such solutions exist they
may not be continuously connected to the naive ones, and their charges
would most likely be different. It is desirable to
find more single soliton solutions which will fill the missing weights in
the fundamental representations (the soliton of species $i$ has topological
charges which are in general a subset of the weights of the fundamental
representation $V_i$ \cite{McGhee}), a problem which
requires urgent attention in the context of the quantum theory and the
exact S-matrices \cite{PRJ}. It is also extremely intriguing that the
weights of the fundamental representations $V_1$ and $V_n$ are filled
for the $A_n$ theories, and it is only for these soliton species $1$ and
$n$ that there
are precisely only a maximum of two parameters in the single soliton
solutions, the case discussed in \S 2.1, 
so that the complication due to the different rates of
rotation of the phases of $U_k$ does not occur, and that the singular
regions in the $Q$-plane are only curves. 
\psbild{ht}{4}{2cm}{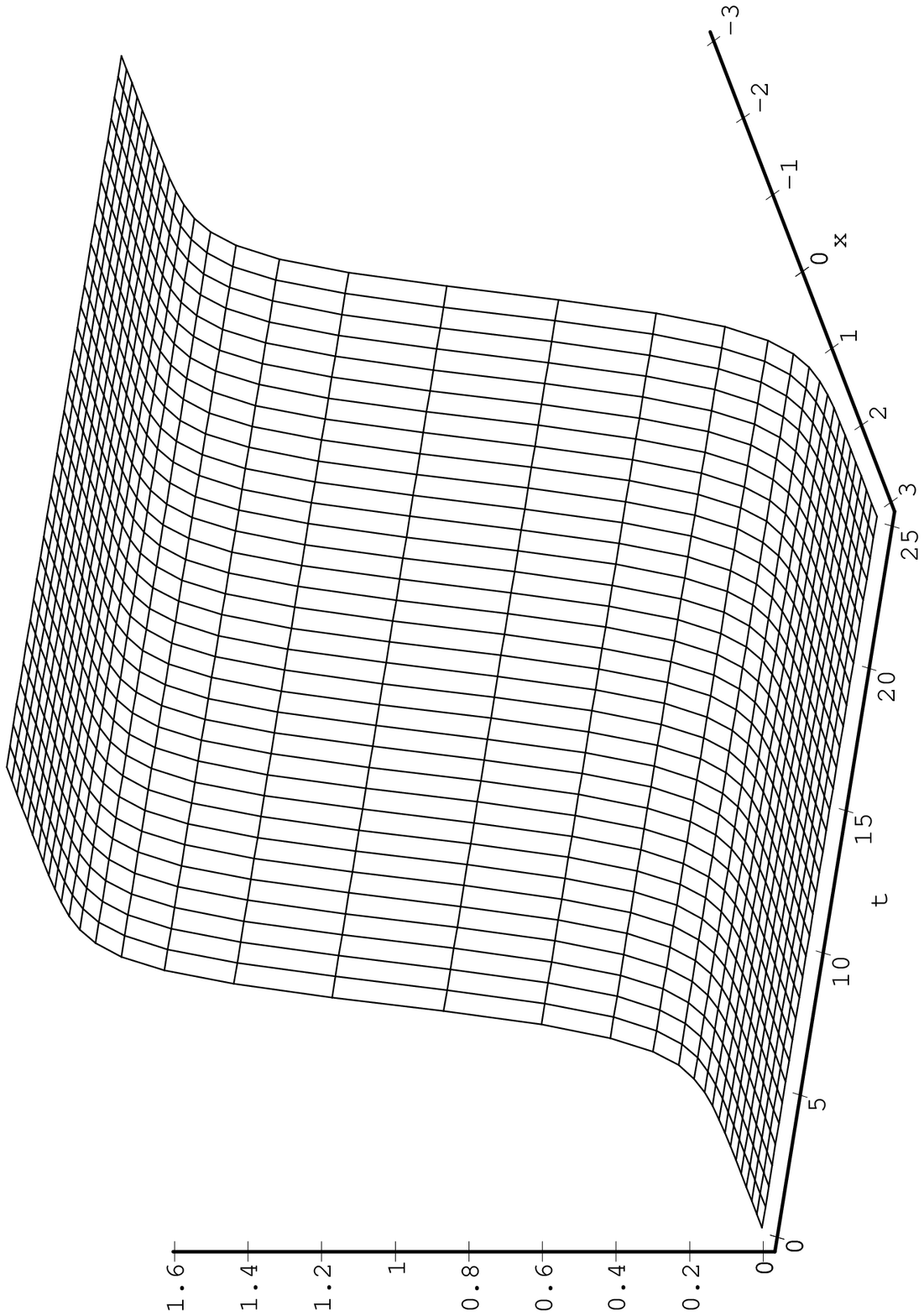}{Plot of ${\rm Re}(\lambda_1\cdot\phi)$ of
equation (\ref{eq: new_sol}) for $Q_1=0$}
\psbild{ht}{5}{2cm}{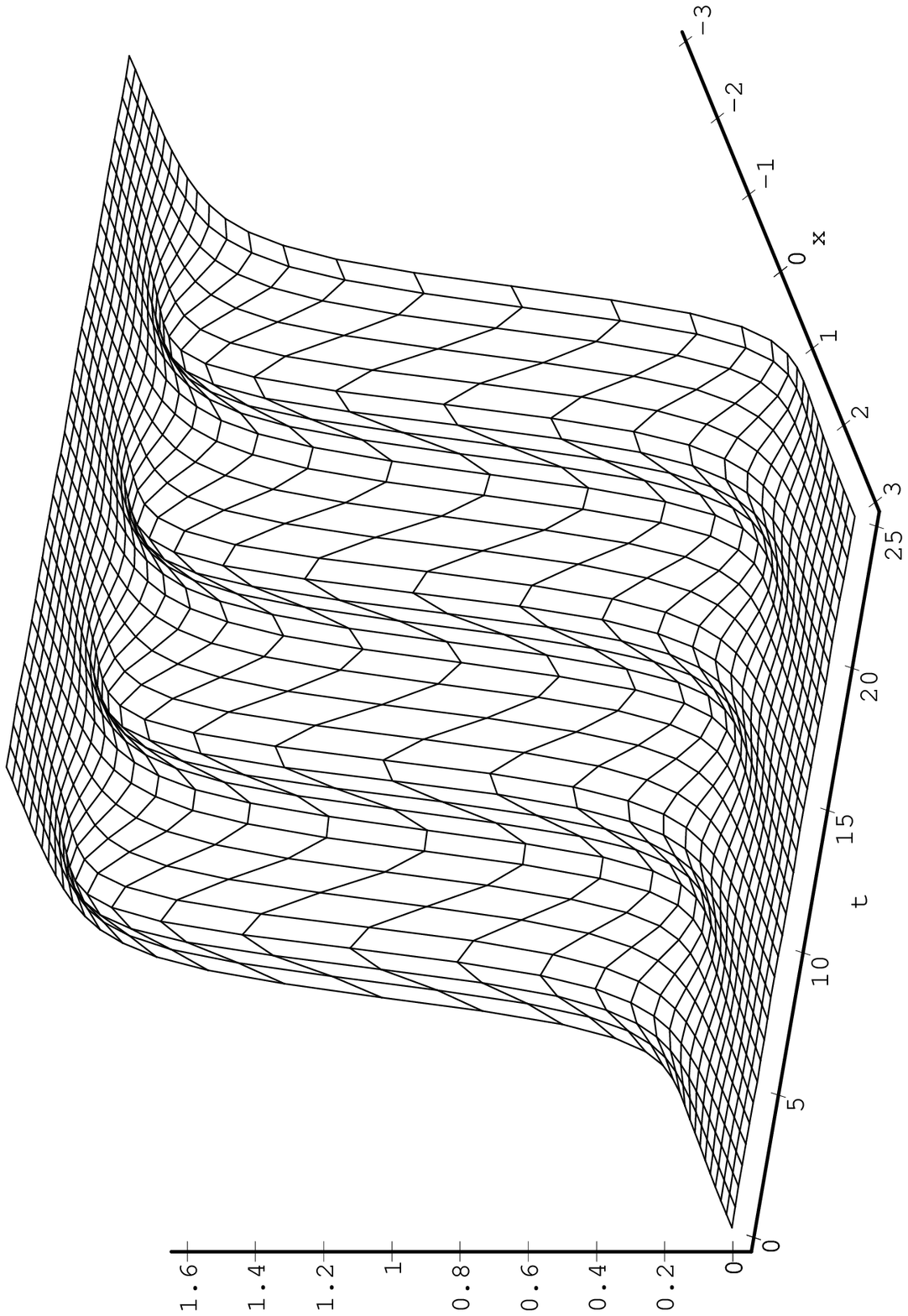}{Plot of ${\rm Re}(\lambda_1\cdot\phi)$ of
equation (\ref{eq: new_sol}) for $Q_1\neq 0$}

\noindent{\bf Acknowledgements}\vspace{0.5cm}

One of us (PRJ) would like to thank Prof. David Olive for introducing
him to the affine Toda solitons, particularly for explaining the
singular regions in the $Q$-plane for the previously known 
single soliton solutions.

\end{document}